# Recognizing Beam Profiles from Silicon Photonics Gratings using Transformer Model

Yu Dian Lim[1,*], Hong Yu Li[2], Simon Chun Kiat Goh[3], Xiangyu Wang[2], Peng Zhao[1,#], and Chuan Seng Tan[1, 2]

[1]*School of Electrical and Electronics Engineering, Nanyang Technological University, 639798, Singapore*
[2]*Institute of Microelectronics, Agency for Science, Technology and Research (A\*STAR), 117685, Singapore*
[3]*Device Solutions Research Singapore, Samsung Electronics, 117440, Singapore*

*[\*]yudian.lim@ntu.edu.sg, [#]ZHAO0275@e.ntu.edu.sg*

**Abstract:** Over the past decade, there has been extensive work in developing integrated silicon photonics (SiPh) gratings for the optical addressing of trapped ion qubits in the ion trap quantum computing community. However, when viewing beam profiles from gratings using infrared (IR) cameras, it is often difficult to determine the corresponding heights where the beam profiles are located. In this work, we developed transformer models to recognize the corresponding height categories of beam profiles of light from SiPh gratings. The model is trained using two techniques: (1) input patches, and (2) input sequence. For model trained with input patches, the model achieved recognition accuracy of 0.938. Meanwhile, model trained with input sequence shows lower accuracy of 0.895. However, when repeating the model-training 150 cycles, model trained with input patches shows inconsistent accuracy ranges between 0.445 to 0.959, while model trained with input sequence exhibit higher accuracy values between 0.789 to 0.936. The obtained outcomes can be expanded to various applications, including auto-focusing of light beam and auto-adjustment of z-axis stage to acquire desired beam profiles.

## 1. Introduction

Free-space optical setups, such as conventional mirrors and lenses, are widely employed in ion trapping system. In a conventional optical setup for ion trapping system, laser beams are guided through the vacuum chamber window to aim the ions, while the fluorescence emitted from the ions is captured by a photomultiplier tube located outside the chamber. However, as the number of trapped ions increases, the optical input and output interfaces used for controlling and measuring individual ions become increasingly constrained [1]. To address these challenges, integrated silicon photonics (SiPh) devices, built on advanced silicon-based microfabrication techniques, have been continuously adopted in ion trap quantum computing [2].

Attributing to its potential for wafer-scale integration with low integration and fabrication costs [3], the development of SiPh has been dynamic. The development of various SiPh devices, such as optical switches [4], micro-ring-resonator [5], and on-chip laser [6] has been extensively-reported. In the context of ion trap integration, single photon detectors are on-chip integrated underneath the ion to facilitate high efficiency florescence collection [7,8]. Similarly, waveguide and grating couplers have been widely used for flexible light routing and multi-location ion addressing [9]. In 2020, various grating couplers for complete wavelength range of $Sr^+$ ion have been integrated to demonstrate the full optical functionality. However, alignment errors in multiple beams present a significant challenge that affects performance [10]. This issue will become even more critical with the introduction of complex operations, such as ion shuttling and multi-qubit operations [11,12].

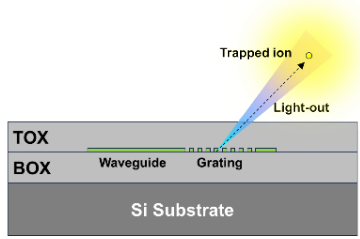

Fig. 1 Illustration of SiPh grating for optical addressing of trapped ion

The illustration of integrated SiPh gratings on the optical addressing of trapped ion is illustrated in Fig. 1. To accurate "aim" the coupled light towards the trapped ions, two key factors should be taken into consideration: (1) the direction of light propagation, and (2) the beam profile morphologies at various heights above the grating. For coupling direction, the angle of the angle of the output light, θ, can be expressed as:

$$\Lambda = \frac{m\lambda}{N - n_1 \sin\theta} \quad (1)$$

where $\Lambda$, m. $\lambda$, N, and $n_1$ are the grating pitch, mode, wavelength, effective index of the grating, and index of the fiber, respectively [13]. On the other hand, as the ion-trapping height scales with the ion trap dimension, it is important to understand the variations in beam profiles at various heights as the light propagates. In a typical beam profiling system in experimental setup, an infrared (IR) camera, attached with a microscopic lens, is usually used to capture the beam profiles coupled out from the gratings (ref. Fig. 2(d)). Conventionally, the distance between the lens and the sample chip is manually adjusted to obtain best-focusing light beam, as done in our previously-reported work [14]. However, such technique does not give us any direct information about the position where the captured beam (on the IR camera) is located. Thus, we attempt to development a machine learning model to recognize the corresponding height of the captured beam profiles.

Among various machine learning models, much work has been reported on the transformer model in various applications. Transformer model was first introduced by Google Inc. as a powerful model for natural language processing (NLP). Due to its self-attention mechanism, transformer model is able to address the attention matrix between individual inputs (or words, in the context of NLP). Thus, it is also one of the key models used in the commonly-known generative artificial intelligence (AI) platform, ChatGPT (GPT - generative pre-trained transformer) [15]. Besides NLP, transformer models can also be used in wide range of applications. In photonics application, our research group has reported application of transformer models in performing next-value prediction on photonics datasets [16,17]. Prediction accuracy of >90% has been achieved.

Leveraging on the success of transformer model, Google Inc. introduces vision transformer (ViT) model. Used in image recognition, an image is first broken down into patches, the patches are inserted into the ViT model where the self-attention layer will address the correlation between individual patches. In the realms of photonics, *Jin et al.* reported proposed a ViT-empowered physics-driven deep neural network which can realize the generation of omnidirectional 3D holograms with small inter-plane crosstalk and high axial resolution [18]. Meanwhile, Merabet *et al.* proposed ViT model in orbital angular momentum mode recognition, which can be used in free-space optical communication [19]. On the other hand, Cuenat *et al.* reported integration of ViT model into digital holographic microscopy for fast autofocusing [20]. However, despite significant presence of ViT models in the abovementioned works, the usage of ViT in SiPh and integrated photonics has been scarce.

In this work, we reported the development of transformer models which can recognize the location of a beam profile from silicon photonics grating, captured on the IR camera. The IR camera is first fixed at a position above the silicon photonics chip as $z = 0$ µm. After that, the IR camera is elevated periodically, where beam profiles were acquired at every 5 µm increment in elevation, in the range of $z = 0 - 805$ µm. The obtained beam profiles are then segregated into Region A, B, C, and D according to the z-position they are located, and then used for training of the transformer models. The recognition accuracies of the trained models are then evaluated, and the model-training processes are repeated 150 times to evaluate the consistency of the obtained outcomes.

## 2. Experimental Methods

The fabrication of gratings is carried out using conventional full-wafer CMOS fabrication process on a 12-inch silicon (Si) substrate, carried out in Institute of Microelectronics (IME), Singapore. The cross-sectional architecture of the sample chip is illustrated in Fig. 1. First, 3 µm buried oxide (BOX) layer is deposited on the Si substrate. Then, 400 nm silicon nitride layer is deposited onto the BOX layer. Photolithography patterning is then carried out on the surface of the silicon nitride layer using immersion lithography technique, followed by dry-etching to etch through the 400 nm silicon nitride layer. At this stage, the morphological structures of the gratings have been formed. Finally, 3 µm top oxide (TOX) layer is deposited onto the etched grating structures.

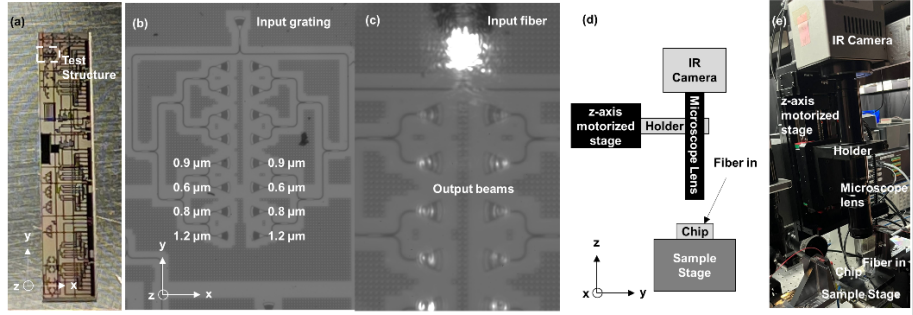

Fig. 2 (a) Sample chip fabricated in IME, (b) Microscopic image of test structure used in this work, (c) output beams from the grating (d) and (e) Equipment used in this work to acquire beam profile datasets

Fig. 2(a) and (b) shows the chip fabricated by the abovementioned process, and the microscopic image of the test structure. The test structure comprises of an input grating connected to 16 output gratings via waveguides and multi-mode interferometers (MMIs). At the same time, the waveguides and the MMIs in the test structure were designed to have equal light energy distributed from the input grating to each output gratings. The pitch of the input grating is 0.8 µm, an optimized pitch to maximize the coupling efficiency of fiber-to-chip coupling of light with 1,092 nm wavelength. The 1,092 nm wavelength is selected as it corresponds to the 'clear-out' function when performing optical addressing on trapped $^{88}Sr^+$ ion qubits [21,22]. Both input and output gratings have dimensions of 30×20 µm along x × y axis, and the duty cycle of all gratings are fixed to be 0.5. From Fig. 2(b), duplicated gratings of 0.6, 0.8, 0.9 and 1.2 µm have also been included in the test structure. This is to increase the number of datasets involve in the model training, where the details to be discussed later.

In a typical measurement, 1,092 nm wavelength light is coupled into the input grating through the input fiber. Then, light propagates from the input grating, along the waveguides and the MMIs, to reach the output gratings. To perform fiber-to-chip alignment onto the grating, first, fiber connected to 1,092 nm laser is aligned towards the input grating under optical microscopic camera. When light beams are visible from the grating (Fig. 2(c)), the optical microscope camera attached to the microscopic lens replaced by an IR camera. The measurement setup to acquire the beam profile data used in this work is illustrated in Fig. 2(d). The beam profiles of light coupled out from the output gratings are then acquired by the infrared (IR) camera. The IR camera used in this work is Ophir XC-130, controlled by BeamGage® Professional software. As mentioned earlier, beam profiles of light from these gratings at various heights are to be acquired. The z-position of the microscope lens is first adjusted to obtain sharpest-possible beam profiles on the IR camera, by adjusting the position of the z-axis motorized stage. The microscope lens consists of a 12× body tube attached to a 2× optical adapter. After obtaining the sharpest beam profiles, the IR camera is moved to ~300 μm lower than the sharpest position. In the context of this work, this position is defined as $z = 0$ μm. From $z = 0$ μm, the microscope lens and the IR camera is gradually moved up, and a beam profile data is taken for every 5 μm it moves up.

For every 5 μm elevation, a beam profile dataset is acquired from the IR camera. The elevation-acquisition process is repeated from $z = 0$ μm to $z = 805$ μm. The beam profile acquisition steps are repeated for output gratings with 0.6, 0.8, 0.9 and 1.2 μm, from both sides of the test structure along x-axis (ref. Fig. 2(b)). The beam profile datasets are acquired and expressed in the form of intensity distributed, with cnt/s as the factory-default unit of each pixel captured by the Ophir XC-130 IR camera and the BeamGage® Professional software. The obtained datasets are then cropped and reshaped for further data analysis and model training, where the details of these will be described in the next section.

## 3. Beam Analysis and Categorization

Fig. 3(a) shows the beam profiles from gratings with various pitches, as viewed under the infrared (IR) camera. The pixel count of the IR camera used in this work is 255×x321, where each pixel represents size of 30×30 μm as a factory-default parameter of the IR camera and BeamGage® Professional software. However, as the microscope lens involves (Fig. 2(d) and (e)), the actual size of each pixel differs. As mentioned earlier, the microscope lens consists of a 12× body tube attached to a 2× optical adapter. Thus, the size of each pixel should be 1.25×1.25 μm after magnification. To further validate this, we used this pixel size to calculate the distance between beam profiles. For instance, the distance between beams from gratings with 1.2 μm and 0.8 μm pitches is 83 pixels. By using the pixel size of 1.25×1.25 μm, the distance between these beams should be ~103.75 μm. Referring back to the microscopic image in Fig. 2(b) and the design layout of the test structure, the center-to-center distance between these two gratings are ~105.5 μm. Thus, it can be deduced that the calculated pixel size of 1.25×1.25 μm is suitable for this work.

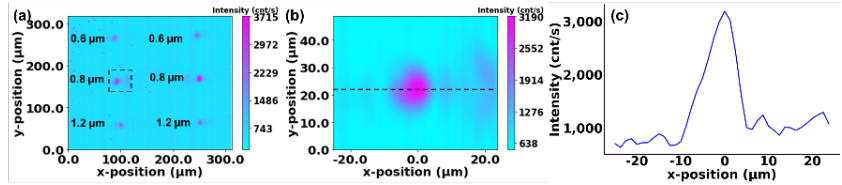

Fig. 3 (a) Beam profiles captured by the IR camera, (b) magnified beam profile of light coupled out from grating with 0.8 μm pitch, (c) intensity distribution along x-axis (beam profiles obtained at $z = 60$ μm)

It can be observed that light beams from gratings with different pitches show different intensity distributions. This aligns with the previous findings, where the optimized pitch of the grating scales with the wavelength of the light used for coupling. The light beam from 0.8 μm shows the highest intensity as compared to its 0.6 μm and 1.2 μm counterparts. This agrees well with the optimization of input grating mentioned in the previous section, where the pitch of the grating was optimized to 0.8 μm for the coupling of 1,092 nm light. From the magnified view of the light beam shown in Fig. 3(b), it can be seen that the beam appeared to be focused, with the presence of multi-mode beams and undesired scattering patterns. To ease the analysis of the intensity distribution along x-axis, we standardized the point where the maximum point of the intensity distribution in the beam is located as x = 0 µm, then we plotted the intensity values ranging between x = ±25 µm of this point. The outcome of the plot is shown in Fig. 3(c). From Fig. 3(c), distinctive peak is observed along x-axis, with the beam width estimated to be ~15 µm. Nevertheless, beam profiles in Fig. 3 was taken at the sharpest-possible z-position. As the microscope lens and the IR camera moves along z-axis, the resulting beam widths may vary.

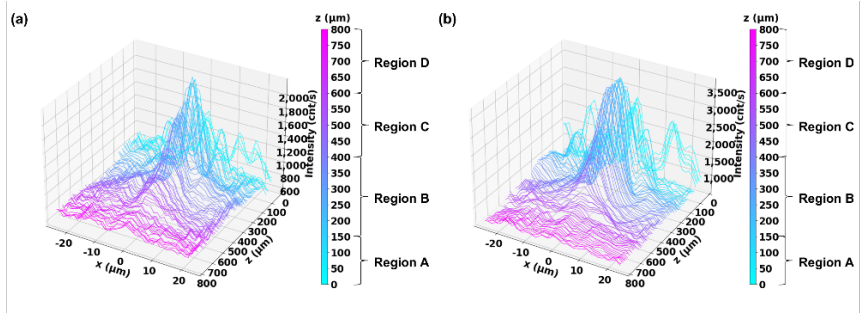

Fig. 4 Intensity distribution along x-axis under various z-position for light beam coupled out from gratings with: (a) 0.6 µm pitch, (b) 0.8 µm pitch

The light intensity distributions along x-axis shown in Fig. 3(c) are plotted at various z-position, as shown in Fig. 4. For light beam from grating with 0.6 µm pitch, the light intensity is relatively lower, where maximum value of ~2000 cnt/s is obtained. On the other hand, light beam from 0.8 µm pitch grating has higher intensity, with maximum value of ~3500 cnt/s. Across z = 0 to 805 µm, we categorized the beam profiles to Region A (z = 0 to 150 µm), Region B (z = 150 to 400 µm), Region C (z = 400 to 600 µm) and Region D (z = 600 to 805 µm), according to the typical morphology of beam profiles in the abovementioned categories.

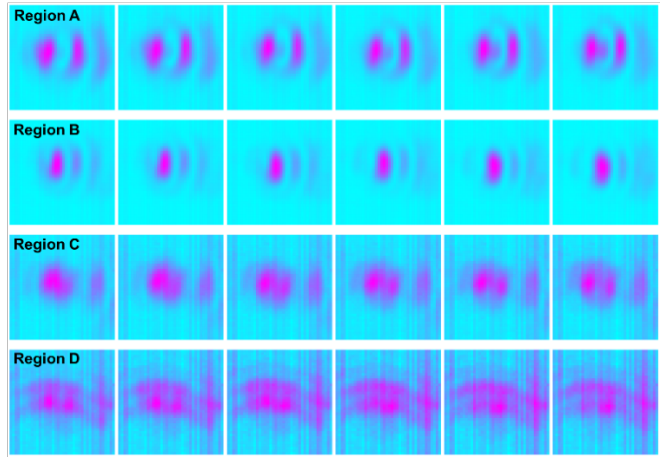

Fig. 5 Typical beam profiles in Region A, B, C and D

The typical morphologies of beam profiles at each region are presented in Fig. 5. In Region A, multiple beam spots are observed, where these spots can be postulated to represent the multiple modes present in this region. At the same time, the beam spots appeared to have curvy morphologies. As this region is the nearest to the grating, the curvy morphologies could be originated from the curved-pitch design in the grating. In Region B, single brightest beam spot with high level of focusing is obtained. The curvy morphologies in Region A are still observable in this region, but less prominent as compared to Region A. At this region, the beam profiles have the best focusing, with narrowest beam width and highest intensity, as observed in Fig. 4 and Fig. 5. Moving further to Region C, the beam profiles became slightly distorted. While the presence of single brightest beam spot is still observable, the intensity of the beam spot reduces significantly as compared to Region B. In Region D, the beam profiles are highly distorted. At the same time, the intensity distribution in the beam is the lowest among the investigated region. As a result, the intensity of the unwanted noises (the vertical lines observed in Region C and Region D) became more prominent, where the noises have relatively similar intensity with the obtained beam profiles.

In developing a transformer model to recognize and distinguish beam profiles from these regions, several preparatory steps should be taken. First, the beam profiles in Region A, B, C and D should be labelled as category '0', '1', '2', and '3', respectively. Then, the beam profiles should be reshaped to standardized dimension prior to inserting to the machine learning model. Third, the beam profiles should be normalized when developing the model. As gratings with different pitches result in beams with different intensities, normalizing the intensity distribution in the beam profiles help the model generalize the beam profiles according to its relative intensity distribution within the profile, instead of the resulting beam intensity from different pitches, at different z-position. The details of the model development and training will be described in the next section.

## 4. Model Development and Evaluation (Patches Input)

Fig. 6 shows the overview of the transformer model. The architecture of the transformer model is based on the vision transformer model that was previously-reported by Google Inc. As mentioned in the previous section, the beam profiles should be reshaped to standardized dimension for the training of transformer model. In this work, we cropped individual beam profile (presented in Fig. 3(b)) from the full-view of the IR camera (presented in Fig. 3(a)). Each beam profile consists of 40×40 pixels, where each pixel has individual cnt/s value. The 40×40 pixels are first divided into patches of 4×4, with each patch comprising of 10×10 pixels. These patches are used as the input of the model training.

Prior to inserting the 4×4 patches dataset, the dataset is first reshaped to (1, 16, 100), which represents 16 patches, and each patch consists of 100-pixel datapoints. The reshaped data is first subjected to positional embedding to embed patches of 1 – 16. Then, it passed through a fully-connected neural network (NN) layer, where the outcome of the NN layer is added to the outcome of the positional embedding. The abovementioned steps are categorized under as patch encoder block. After that, the outcome of the patch encoder will undergo a series of normalization layers, NN layers, and self-attention layer in the transformer encoder block, as shown in Fig. 6. For self-attention layer, the attention correlations between each and every patch of the 16 patches inserted into the layer. The details of the self-attention mechanism are described in ref. [15,23]. After the transformer encoder block, the outcome of the block will be flattened from (1, 16, 100) to a series of 1600 datapoints. Then, it passes through two layers of NN layer to reach the final output recognition.

For each NN layer, the number of nodes per layer is fixed at 64. For the self-attention layer, we used a single-head attention layer with head size of 16. The number of epochs is fixed at 100. However, an early-stopping condition is applied. If the validation accuracy value did not improve for 30 consecutive epochs, the training will stop, and the model will be restored to the epoch with highest validation accuracy. The details of the transformer model is given in the full Python code in ref. [24].

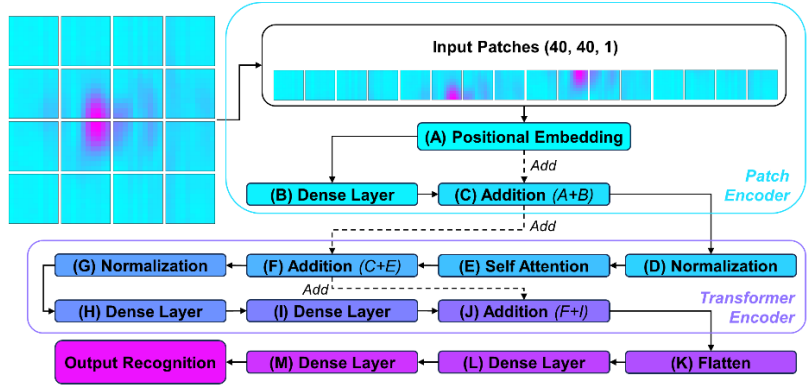

Fig. 6 Overview of the transformer model

In this work, a beam profile dataset (presented in Fig. 3(b)) is captured and filtered as 40×40 pixel, every 5 μm elevation from $z = 0$ to 805 μm. Thus, 161 beam profile datasets are captured in this range. Light from gratings with pitches of 0.6, 0.8, 0.9 and 1.2 μm are investigated and cropped from the full view of the IR camera (presented in Fig. 3(a)), where the profiles of each grating are acquired twice, from duplicated gratings (ref. Fig. 2(b). Thus, we have 4×2×161 = 1,288 beam profile datasets ranging from $z = 0$ to 805 μm. For the training of transformer model, 60% (772) of randomly-split datasets will be used, where the remaining 516 datasets will be used as the testing datasets to evaluate the performance of the trained transformer model. Among the 772 datasets, 30% of randomly-selected datasets will be used as the validation datasets. The details of the beam profile cropping, data compilation, data splitting (training, validation, and testing datasets) are given in the full Python code in ref. [24].

As mentioned earlier, the transformer model will stop training if the validation accuracy did not improve for 30 consecutive epochs, and the model will restore to the epoch with highest validation accuracy. After that, the testing beam profile datasets will be inserted to the model to evaluate the recognition performance. For instance, the beam profile presented in Fig. 6 are taken at $z = 300$ μm (Region B) from grating with 0.6 μm pitch. The beam profile is first broken into patches, as presented in Fig. 6. The patched profile is then reshaped to (1, 16, 100), as stated in Fig. 6. After the patch encoder, the dimension of the dataset remained to be (1, 16, 100), as shown in Fig. 7(a). The values of the dataset ranges between 600 to 2000, which represents the cnt/s values of each pixel in the 40×40 dataset from the grating with 0.6 μm pitch (ref. Fig. 4(a)). After that, the dataset will undergo a series of processes in the transformer encoder block. After the transformer encoder block, the range of values converted to -1000 to 1500 but the dimension remained at (1, 16, 100), as shown in Fig. 7(b). Then, the dataset is flattened (layer K in Fig. 6) to a series of 1600 datapoints (Fig. 7(c)), and passes through two layers of full-connected neural network (NN) layer before reaching the profile recognition step. At the output recognition step, the transformer model computed the given beam profile as 'Category 1', which corresponds to Region B. Since the beam profile was taken $z = 300$ μm from grating with 0.6 μm pitch, the trained model made a correct recognition.

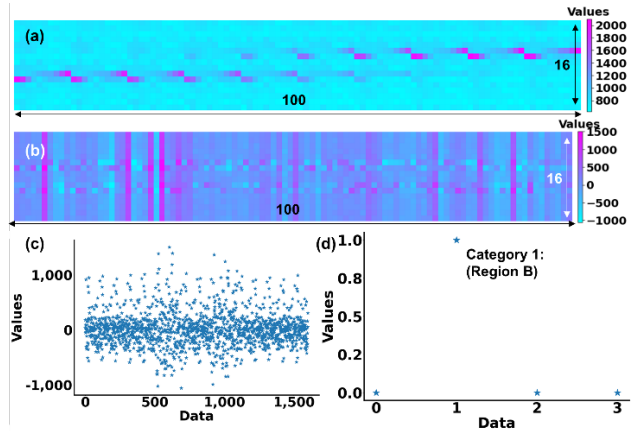

Fig. 7 (a) Dataset after patch encoder block, (b) dataset after transformer encoder block, (c) dataset after flattening (layer K Fig. 6), (d) categorization of beam profile

The abovementioned recognition process was repeated for all 516 testing beam profiles, the outcome of the profile recognition is given in Fig. 8. It can be observed that the trained model correctly-recognized almost all 516 testing beam profiles. All 'mistakes' occurs under the circumstance where the model wrongly-recognized a given profile as falling under the neighboring regions. For instance, among the 145 profiles in Region C, the model wrongly recognized 5 beam profiles as Region B and 6 beam profiles as Region D. The model did not mistakenly recognize any beam profiles in Region C as Region A. Similarly, among the 149 profiles in Region B, the model wrongly recognized 2 beam profiles as Region A and 6 beam profiles as Region C. These mistakes can be due to similar morphologies in beam profiles at the edges of each region. For instance, the beam profiles in Region B and Region C ranges between z = 150 – 400 μm and z = 400 – 600 μm, respectively. The beam profiles around z = 400 μm may have overlapped morphological features between Region B and Region C. Nevertheless, the trained transformer model obtained a satisfactory recognition accuracy of 0.938 for 516 testing beam profiles across 4 categories.

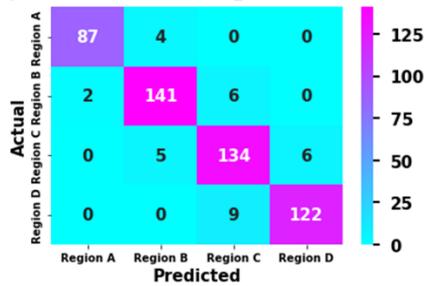

Fig. 8 Confusion matrix of recognizing 516 beam profiles with patch inputs

## 5. Model Development and Evaluation (Sequence Input)

The training and evaluation of transformer model in the previous section was done by inserting beam profiles with 40×40 pixels. However, observing the changes in beam profile morphologies across z = 0 to 805 μm, the prominent changes across different heights are the peak intensity values, and the intensity distribution along x-axis (ref. Fig. 4). However, the peak intensity values do not sufficiently represent its corresponding region. For instance, for a beam profile with peak intensity of ~2000 cnt/s, it is hard to determine whether it falls at Region B of light from 0.6 μm pitch grating, or Region A or C of light from 0.8 μm pitch grating. An alternative, and possibly faster way of training the transformer model for beam profile recognition is to use a sequence of cnt/s values along x-axis. This means that we use the intensity plots along x-axis illustrated in Fig. 4 as the input for training the transformer models.

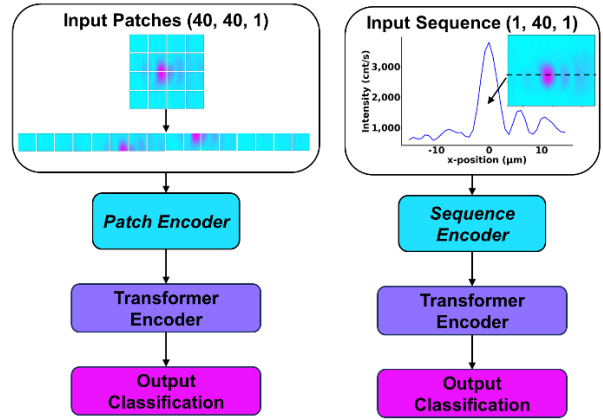

Fig. 9 Comparison between training transformer model using input patches and input sequence

Fig. 9 compares the methods of training transformer model using input patches and input sequence. As mentioned earlier, each beam profiles are cropped into dimension of 40×40 pixels. The process of training the transformer model using input patches has been described earlier. For training using input sequence, we breakdown the 40×40 pixels beam profiles into a sequence of 40-pixel values along x-axis. To extract the 40-pixel sequence from the beam profiles, first, the position with the highest cnt/s pixel in the beam profile is located. This position should be presumably the peak of the beam. Then, the previous 20 pixel-values, and the next 19 pixel-values along the x-axis is taken as a sequence. From these steps, we obtain a sequence with 40 pixel-values, with the maximum point located at the 21$^{st}$ position in the sequence. For the training and recognition and beam profiles, the trained transformer model will use the 40 pixel-value sequence, instead of the patches described in Fig. 6.

The abovementioned profile-to-sequence process is repeated for all 1,288 acquired beam profiles. Similar to the previous section, the 1,288 sequences are split into 60% (772) training datasets and 516 testing datasets. Among the 772 training datasets, 30% of randomly-selected datasets will be used as the validation datasets. The datasets are fed into the transformer model for training and evaluation purposes. The parameters of each layer used in the transformer model, and the number of epochs, stopping condition of training, and all other training conditions remained identical.

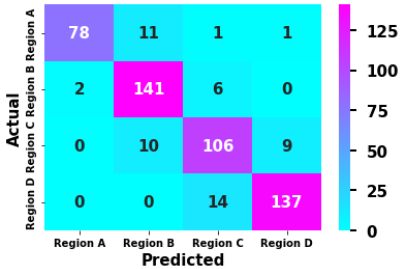

Fig. 10 Confusion matrix of recognizing 516 beam profiles with sequence inputs

Fig. 10 shows the confusion matrix when recognizing 516 testing beam profiles with sequence inputs. As compared to recognizing with patch inputs, recognizing with sequence input yields slightly lower accuracy. For instance, when recognizing beam profiles from Region

C, the model made more mistakes. Among the 125 beam profiles in Region C, the model wrongly-recognized 10 beams as Region B, and 9 beams as Region D. Do note that the train-test splitting of beam profile datasets are carried out randomly. Thus, the number of beam profiles present in each category is not the same for Fig. 8 and Fig. 10. Nevertheless, the total number of testing beam profiles remained to be 516 for both Fig. 8 and Fig. 10. It is clear that the transformer model trained using input sequence is slightly underperformed as compared to model trained using input patches, with lower accuracy of 0.895 when recognizing 516 beam profiles. This can be attributed to the presence of less representative datapoints, such as the pixel values that illustrates the overall beam profile morphologies. The details of extracting 40-pixel values data, data compilation, data splitting (training, validation, and testing datasets) are given in the full Python code in ref. [25].

Despite having lower accuracy, transformer model trained using input sequence can possibly offer a faster, more stable alternative as compared to input patches. The comparison between transformer models trained with both methods will be discussed in the next section.

## 6.  Repeatability Testing

Fundamentally, the training of machine learning models refers to the optimization of internal parameters in the model. In the context of this work, each layer in a transformer model illustrated in Fig. 6 has a series of parameters. Thus, training datasets with inputs (patches or sequences) and outputs (categorization of beam profiles) are used to optimize parameters in the model to suit the datasets. The process of training a model can be probabilistic. This means that, for each training, the parameters in the model, and the associated predictive ability of the model, can be varied.

As mentioned in the previous section, 1,288 beam profiles are split into training, validation, and testing datasets. The splitting itself is carried out randomly, and the model training is probabilistic. To test the robustness of our work, we repeated the dataset splitting, model training, and model evaluation for 150 times, using both input patches and input sequences as the input datasets. The accuracy values in recognizing the 516 testing beam profiles are recorded. At the same time, training loss and validation loss of the trained model at the best epoch (lowest validation accuracy) are also recorded. To determine whether the model is well-fitted to the dataset, the loss-difference values (training loss-validation loss) are calculated for all 150 runs. The outcomes are presented in Fig. 11.

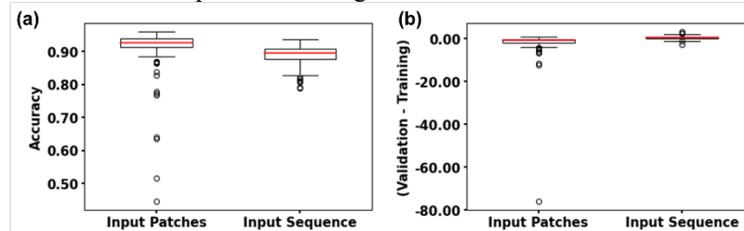

Fig. 11 Boxplots from 150 model trainings using input patches and input sequence: (a) evaluate accuracies, (b) training loss-validation loss

From Fig. 11(a), it can be seen that most the overall accuracy is higher when the transformer model is trained using input patches. The range between the minimum and maximum points of the box plot is 0.886 to 0.959. Meanwhile, for model trained using input sequence, the range is 0.828 to 0.936. Despite having wider range in boxplot, transformer model trained with input sequence has less outliers. For model trained with input patches, 14 outliers that range between from 0.445 to 0.870 are present in the list of 150 accuracy values. On the other hand, upon repeated training, transformer model trained with input sequence only have 8 outliers, ranging from 0.789 to 0.819. It can be postulated that transformer trained by input patches are less stable in terms of beam-recognition performance, due to underfitting of the model. This postulation is reflected in Fig. 11(b), where the loss-difference values for model trained by input patches are more negative, where the maximum and minimum values in the boxplots are 1.058 and -3.759, respectively. On the other hand, the loss-difference values for model trained by input sequence have maximum/minimum values of 2.243/-1.028 in within the boxplot. The lowest outliers from 150 repetitions when using input sequence is -2.547. Meanwhile, the lowest outliers when using input patches is -76.08, with several loss-difference values ranging between -5.43 to -12.1.

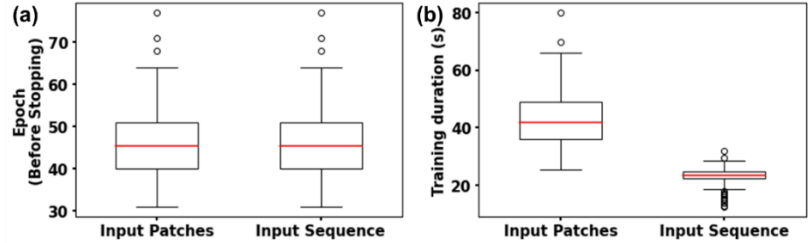

Fig. 12 Boxplots from 150 model trainings using input patches and input sequence: (a) Number of Epoch before stopping, (b) training duration

Besides recognition accuracy and loss-difference values (training loss-validation loss), another key figure-of-merit to compare models trained by input patches and input sequence is the training time. As mentioned in the previous section, if the validation accuracy value did not improve for 30 consecutive epochs, the training will stop, and the model will be restored to the epoch with highest validation accuracy. As model training is probabilistic, each round of training may have different training and validation accuracies. Thus, for each training cycle, the model training is ought to stop at different epoch. Despite so, from Fig. 12, it can be seen that the distributions of total epoch are similar for models trained with input patches and input sequence. Under this condition, transformer model trained using input sequence took shorter training time, as compared to its input patches counterpart. This can be attributed to the presence of smaller-size datasets when training with input sequence. For input patches, each profile with 40×40 pixels are reshaped and inserted into the model; meanwhile, for input sequence, a sequence of 40 pixels is extracted from each profile to feed to the model. The full Python code (with linked training loss, validation loss, evaluated accuracies, confusion matrix, etc.) to plot Fig. 11 and Fig. 12 is given in ref. [26].

## 7. Discussion

The fundamental concept of the transformer model described in Fig. 6 is based on the vision transformer (ViT) model introduced by Google Inc. used for image recognition. Typically, the image will be reshaped to a suitable dimension, say 40×40 pixels. For 40×40 dimension, the image is first split into 4×4 patches, with each patch having 10×10 pixels. For each pixel, it contains red, green, blue (RGB) color codes that range between 0 to 255. Thus, prior to training the ViT model with the image, the images are first converted to individual datasets of (40, 40, 3), which further reshaped to (1, 16, 300) after the patch encoder and transformer encoder. Within (1, 16, 300), 16 represents the number of patches, and 300 represents number of datapoints (100 pixels × 3 RGB values) in each patch. Benchmarking against this work, we used similar ViT model, but we make the input beam profiles to have dimension of (40, 40, 1). Instead of having RGB color codes that range between 0 to 255, in each pixel, light intensity in expressed in cnt/s is present. Thus, input data with dimension of (1, 16, 100) is used. As a result, the number of datapoints inserted into the model is 3 times smaller, which significantly reduces the training time of the model.

In the transformer encoder block illustrated in Fig. 6, the self-attention layer (layer E) is the key component in the transformer model. In the context of this work, it computes individual attention coefficient of individual patches. For instance, if we encode the patches as "A, B, C,

D... etc.", the self-attention layer computes attention coefficients of "A-A, A-B, A-C, A-D, B-C, B-D, C-D... etc.". As a result, the relationship between individual patches is addressed, and the transformer model can better understand the correlation between individual patches and the categorization of the beam profile formed by the patches. Similarly, if we input a sequence with 40-pixel values instead of patched profile dataset with 40×40-pixel values, the attention coefficients between individual elements in the 40-pixel values will be computed.

Comparing techniques of training the transformer models using input patches and input sequence, when using input patches, the attention coefficients between every pair of patches are addressed. However, some patches may be irrelevant to the beam profile. Taking the patched beam profile presented in Fig. 6 as an example, the patches at the four corners of the profile is irrelevant to the beam. Thus, it may introduce unwanted noise when training the model. On the other hand, when training the transformer model with input sequence, almost all values in the 40-pixel sequence are relevant to the identification of the beam profile according to its position along z-axis (Region A, B, C or D). By performing self-attention step on the input sequence, the model may have missed the additional information of the beam profiles along y-axis, but it introduces less noise. Thus, despite having slightly lower prediction accuracy, transformer models trained with input sequence shows more consistent outcome, as shown in Fig. 11(a). At the same time, the difference between training and validation losses is near zero, indicating that the model is well-fitted.

A significant drawback of training the transformer model with input sequence is the inflexibility. As mentioned earlier, to extract the 40-pixel sequence from the beam profiles, first, the position with the highest cnt/s pixel in the beam profile is located. Then, the previous 20 pixel-values, and the next 19 pixel-values along the x-axis is taken as a sequence. At high z-position like Region D, the pixel with highest cnt/s may not locate at the maximum point of the beam. As presented in Fig. 5, in Region D, the noises (vertical lines) in the profile may have higher pixel values than the beam. To resolve this, the vertical lines need to be manually neutralized to ensure that we extract the correct 40-pixel sequence from every beam profile.

In summary, transformer model trained with input patches has higher accuracy, but the performance is relatively inconsistent. Whereas, transformer model trained with input sequence has slightly lower accuracy, but its performance is more consistent upon training. The selection of model-training technique depends on the requirement where the models to be deployed. If high accuracy is needed, the model can be trained several times with input patches, where model with the highest prediction can be selected. However, the trained model may not perform as good when recognizing wider scope of unseen, newly-added beam profiles. If the prediction accuracy can be slightly compromised, but consistent prediction outcomes are crucial, the model should be trained with input sequence. However, additional steps of neutralizing the noise in the IR camera view is needed.

To facilitate further expansion of this work, we have included the full Python code used in this work for possible applications in wider range of photonics domain. For transformer model trained by input patches, the code is given in ref. [24]; meanwhile, for transformer model trained by input sequence, the code is given in ref. [25]. The full training and evaluation results from the 150 training cycles can be found in ref. [26].

## 8. Conclusion

In this study, transformer models are developed to recognize the corresponding heights of light beams coupled out from the silicon photonics gratings. To obtain the training data, a z-axis motorized stage was integrated with an infrared (IR) camera to acquire beam profiles at precise intervals along z-axis. Beam profiles were acquired at every 5 μm increment in elevation, in the range of z = 0 - 805 μm. The obtained beam profiles are then categorized into Region A, B, C, and D according to their corresponding heights, and randomly split into training, validation, and testing datasets. Two training data types are attempted: (1) beam profiles with 40×40-pixel values, and (2) intensity plot along x-axis with 40-pixel values, named as 'input patches' and

'input sequence', respectively. When recognizing corresponding heights of 516 beam profiles, model trained using input patches achieved accuracy of 0.938; meanwhile, model trained using input sequence achieved lower accuracy of 0.895. When repeating the beam profile splitting, model training, and model evaluation steps for 150 cycles, it is found that model trained using input sequence shows better consistency where accuracy values between 0.789 to 0.936 are obtained. Whereas, model trained by input patches has accuracy values between 0.445 to 0.959. At the same time, transformer model trained by input sequence is better-fitted to both training and validation datasets. Training the transformer model with input sequence also took shorter training time of ~20s, as compared to input patches that took ~40s on average. The obtained outcomes from this work established a foundation to recognize light beam profiles according to their corresponding heights. Upon deploying the trained model to integrate with the IR camera, it can be expanded to various applications, including auto-focusing of light beam and auto-adjustment of z-axis stage to acquire desired beam profiles.

## 9. Back matter

**Funding.** Ministry of Education of Singapore AcRF Tier 2 (T2EP50121-0002 (MOE-000180-01)) and AcRF Tier 1 (RG135/23, RT3/23); National Centre for Advanced Integrated Photonics (NCAIP) (NRF-MSG-2023-0002); National Research Foundation, Singapore, and A*STAR under its Quantum Engineering Program (NRF2021-QEP2-03-P07) and A*STAR SPF (C222517002)

**Acknowledgments.** This work was supported by the Ministry of Education of Singapore AcRF Tier 2 (T2EP50121-0002 (MOE-000180-01)) and AcRF Tier 1 (RG135/23, RT3/23); National Centre for Advanced Integrated Photonics (NCAIP) (NRF-MSG-2023-0002); National Research Foundation, Singapore, and A*STAR under its Quantum Engineering Program (NRF2021-QEP2-03-P07) and A*STAR SPF (C222517002). The preparation of Python codes in ref. [24–26] is partly assisted by the generative AI tool, ChatGPT.

**Disclosures.** The authors declare no conflicts of interest.

**Data availability.** Datasets obtained from IR camera shown in Fig. 3, Fig. 4, and Fig. 5 are available upon request. The full training and evaluation results from the 150 training cycles can be found in ref. [26].